\documentclass[prd,aps,epsfig]{revtex4} 
\usepackage{epsfig} 
\usepackage{subfigure} 
\usepackage{amsmath} 
\usepackage[dvips]{color} 
\usepackage[percent]{overpic}

\def\beq{\begin{equation}} 
\def\enq{\end{equation}} 
\def\beqa{\begin{eqnarray}} 
\def\enqa{\end{eqnarray}}

\def\G3{\lag g^3G^3\rag}

\def\nn{\nonumber}

\newcommand{\rag}{\rangle} 
\newcommand{\lag}{\langle}

\newcommand{\DZero}{\rm D\O}%\emptyset}}

\newcommand{\jpsi}{\ensuremath{J/\psi}}
\newcommand{\psip}{\ensuremath{\psi(2S)}}

\def\d3pi{$D^+\rightarrow \pi^-\pi^+\pi^+$\ }
\def\ds3pi{$D_s^+\rightarrow \pi^-\pi^+\pi^+$\ }

\begin{document} 
 
\title{\sc Phase motion in the $Z^-(4430)$ amplitude in $B^0\to\psi^\prime\pi^-K^+$ decay} 
 
%\title{\sc Phase motion in the $J/\psi(\psi^\prime)\pi^-$ amplitude in $B^0\to J/
%\psi(\psi^\prime)\pi^-K^+$ decay} 

\author{Ignacio Bediaga, Jussara M. de Miranda, Fernando Rodrigues}
%\email{bediaga@cbpf.br}
%\email{jussara@cbpf.br}
%\email{felufero@gmail.com}
\affiliation{Centro Brasileiro de Pesquisas F\'\i sicas, Rua Xavier 
Sigaud 150, 22290-180 Rio de Janeiro, RJ, Brazil}
\author{Marina Nielsen} 
%\email{mnielsen@if.usp.br}   
\affiliation{Instituto   de  F\'{\i}sica, 
  Universidade de S\~{a}o Paulo, C.P.  66318, 05314-970 S\~{a}o Paulo, 
  SP,    Brazil} 
\begin{abstract} 
In view of the proliferation in the number of new charmonium states, it is really 
important to have a experimental way to prove that an observed bump is, indeed, a 
real resonance. To do that, in this paper we present an alternative method to 
demonstrate the resonant behavior of a state. With this method, the  phase variation 
of a generic complex amplitude can be directly revealed through  interference  in 
the Dalitz-plot region where it crosses  a well established  resonant state, 
used as a probe. We have tested the method for the $Z^-(4430)$ state by generating 
Monte Carlo samples for the $B^0\to \psip \pi^-K^+$ decay channel. We have shown that
the proposed method gives a clear oscillation behavior, related to the phase 
variation associated to a real resonant state, in the case where the 
$Z^-(4430)$ is considered as a regular resonance with a strong phase variation. 
We have also discussed the possibility of using the proposed method complementary to 
the Argand diagram to determine the internal structure of the $Z^-(4430)$ state.
   
\end{abstract}  
 
\maketitle 
 
%%%%%%%%%%%%%%%%%%%%%%%%%%%%%%%%%%%%%%%%%%%%%%%%%%%%%%%%%%%%%%%%%%%%% 
% 
% 
%\section{Introduction} 
%  
%  
%%%%%%%%%%%%%%%%%%%%%%%%%%%%%%%%%%%%%%%%%%%%%%%%%%%%%%%%%%%%%%%%%%%%% 

Several experiments operating during the last decade, mainly BaBar 
at SLAC and Belle at KEK, CLEO-III and CLEO-c at CESR, 
CDF and \DZero\ at Fermilab,  BESIII at IHEP and LHCb and CMS at CERN, 
have vastly increased the available data on new charmonium-like states, called 
$X,~Y$ and $Z$ states. Among these states, the charged are the most
interesting ones, since they can not be simple $c\bar{c}$ states. 
The $Z^+(4430)$, found by Belle 
Collaboration in 2007, was the first observed one 
\cite{Choi:2007wga,Mizuk:2009da,bellez2}. 
Since the minimal quark content of this state is $c \bar{c} u \bar{d}$  this  
can only be achieved in a multiquark configuration.  
The BaBar Collaboration searched for the $Z^-(4430)$ signature in
four decay modes and concluded that there is no significant evidence for a 
signal peak  in any of these processes \cite{babarz}. However, very recently 
the Belle  and LHCb collaborations have confirmed the $Z^+(4430)$ observation 
and have determined the  preferred assignment of the quantum numbers to be 
$J^{P} = 1^{+}$ \cite{bellez2,Aaij:2014jqa}.    
Curiously, the first evidence of this resonance in the $J/\psi \pi^+$ channel
was reported only this year by Belle Collaboration \cite{Chilikin:2014bkk}. 

The $Z^+(4430)$ observation motivated further studies of other $\bar{B}^0 $ 
decays and, in  2008, Belle Collaboration reported the observation of other 
two resonance-like structures, called  $Z_1^+(4050)$ and $Z_2^+(4250)$, 
in the exclusive process  $\bar{B}^0\to K^-\pi^+\chi_{c1}$, in the $\pi^+
\chi_{c1}$ mass distribution ~\cite{Mizuk:2008me}.
Once again the BaBar colaboration did not confirm these observations 
\cite{Lees:2011ik}.

Following these observations, from March to October of 2013 four more 
charmonium charged states were reported. 
The first one was the $Z_c^+(3900)$, observed almost at the same time by 
BESIII \cite{Ablikim:2013mio} and Belle\cite{Liu:2013dau} collaborations, in 
the $M(\pi^\pm J/\psi)$ mass spectrum of the $Y(4260)\to J/\psi\pi^+\pi^-$ 
decay channel. This structure was also  confirmed by the  authors of 
ref.~\cite{Xiao:2013iha} using  CLEO-c data.  

Soon after the $Z_c^+(3900)$ observation, the BESIII related the observation 
of other three charges states: $Z_c^+(4025)$ \cite{Ablikim:2013emm}, 
$Z_c^+(4020)$ \cite{Ablikim:2013wzq}    and $Z_c^+(3885)$ 
\cite{Ablikim:2013xfr}.
Up to now it is not clear if the states $Z_c^+(3900)$-$Z_c^+(3885)$
and the states $Z_c^+(4025)$-$Z_c^+(4020)$ are the same states seen in 
different decay channels, or if they are independent states. 

Finally, in August 2014 the $Z_c^+(4200)$ was reported by Belle Collaboration
in the $\jpsi\pi^+$ channel of the $\bar{B}^0$ decay, with a 6.2$\sigma$
significance. As in the case of the $Z^+(4430)$, the prefered assignment
of the quantum numbers is $J^P=1^+$ \cite{Chilikin:2014bkk}.
We show these states in Table~\ref{Spec}.
For more details we refer the reader to the more comprehensive review articles
\cite{Nielsen:2009uh,Brambilla:2010cs,Nielsen:2014mva,Liu:2013waa,Brambilla:2014jmp}.

\begin{table*}[h]
\caption{The new charged states in the $c\bar{c}$ regions, 
ordered by mass. Masses $m$ and widths $\Gamma$ represent
the weighted averages from the listed sources as in 
\cite{Brambilla:2014jmp}, or are taken from \cite{pdg}
when available. The citation given in {\color{red} red} is for the first
observation and the citation given in {\color{blue} blue} is for a non
confirmation. 
The Status column NC (neds confirmation) indicates that the state has been 
observed by only one, or was not confirmed by other experiment. The Status is 
OK if at least two independent experiments saw the state.}
\setlength{\tabcolsep}{0.23pc}
\label{Spec}
\begin{center}
\begin{tabular}{lccclccc}
\hline\hline
\rule[10pt]{-1mm}{0mm}
 State & $m$~(MeV) & $\Gamma$~(MeV) & $J^{P(C)}$ & \ \ \ \ Process~(mode) & 
     Experiment & Year & Status \\[0.7mm]
\hline
\rule[10pt]{-1mm}{0mm}
$Z_c^+(3885)$ & $3883.9\pm4.5$ & $25\pm12$ & $1^{+}$&
     $ Y(4260)\to \pi^-(D\bar{D}^{*+})$ &
     {\color{red} BESIII}~\cite{Ablikim:2013xfr}& 2013 & NC\\[1.89mm]
%&&&&&& \\
$Z_c^+(3900)$ & $3896.7\pm7.3$ & $55\pm35$ & $1^{+}$&
     $ Y(4260)\to \pi^-(\pi^+\jpsi)$ &
     {\color{red} BESIII}~\cite{Ablikim:2013mio}, Belle~\cite{Liu:2013dau}, CLEO-c \cite{Xiao:2013iha}]& 2013 & OK\\[1.89mm]
%&&&&&& \\
$Z_c^+(4020)$ & $4022.9\pm2.8$ & $7.9\pm3.7$ & . . .&
     $ e^+e^-\to \pi^-(\pi^+h_c)$ &
     {\color{red} BESIII}~\cite{Ablikim:2013wzq} & 2013 &NC \\[1.89mm]
%&&&&&& \\
$Z_c^+(4025)$ & $4026.9\pm4.5$ & $24.8\pm9.5$ & $1^+,~2^+$&
     $ Y(4260)\to \pi^-(D^*\bar{D}^*)^+$ &
     {\color{red} BESIII}~\cite{Ablikim:2013emm} & 2013 &NC \\[1.89mm]
%&&&&&& \\
$Z_1^+(4050)$ & $4051^{+24}_{-43}$ & $82^{+51}_{-55}$ & . . . &
     $ B\to K (\pi^+\chi_{c1}(1P))$ &
     {\color{red} Belle}~\cite{Mizuk:2008me}, {\color{blue} BaBar}~\cite{Lees:2011ik}& 2008 & NC\\[1.89mm]
%&&&&&&\\
$Z_c^+(4200)$ & $4196^{+35}_{-30}$ & $370^{+99}_{-110}$ & $1^{+}$&
     $B\to K (\pi^+\jpsi)$ &
     {\color{red} Belle}~\cite{Chilikin:2014bkk}
     & 2014 & NC\\[1.89mm]
%&&&&&&\\
$Z_2^+(4250)$ & $4248^{+185}_{-\ 45}$ &
     177$^{+321}_{-\ 72}$ & . . . &
     $ B\to K (\pi^+\chi_{c1}(1P))$ &
     {\color{red} Belle}~\cite{Mizuk:2008me}, {\color{blue} BaBar}~\cite{Lees:2011ik}  & 2008 &NC\\[1.89mm]
%&&&&&&\\
$Z^+(4430)$ & $4458^\pm15$ & $166^{+37}_{-\ 32}$ & $1^{+}$&
     $B\to K^- (\pi^+\psi(2S))$ &
     {\color{red} Belle}~\cite{Choi:2007wga,Mizuk:2009da,bellez2}, 
{\color{blue} BaBar}~\cite{babarz}, LHCb~\cite{Aaij:2014jqa}
     & 2007 & {OK}\\[0.7mm]
& & & & $B\to K^-(\pi^+ J/\psi)$ &
    Belle~\cite{Chilikin:2014bkk} & &\\[1.89mm]
%&&&&&&\\
\hline\hline
\end{tabular}
\end{center}
\end{table*}
 
In view of so many non confirmed (NC) states in Table~\ref{Spec}, it is really 
important to have
a experimental way to prove that an observed excess of events is, indeed, a real 
resonance. In particular, bumps close to the threshold of a pair of particles 
should be treated with caution \cite{Aceti:2014uea}. Sometimes they are identified 
as new particles,
but they can also be a reflection of a resonance below threshold. As an 
example, in the case of the $Z_c^+(3885)$, it was shown in 
Ref.~\cite{Aceti:2014uea} that the signal reported in \cite{Ablikim:2013xfr} 
could be also described by a $D\bar{D}^*$ resonance with a mass around 
3875 MeV and width around 30 MeV. Also, in the case of the $Z_c^+(4025)$, it 
was shown in Ref.~\cite{Torres:2013lka} that both, a resonance with $J^P=1^+$ 
or a bound state with  $J^P=2^+$, are compatible with the data from
Ref.~\cite{Ablikim:2013emm}. Besides, it was also shown
in ref.~\cite{Torres:2013lka} that the experimental data can also be
explained with just a pure wave-D background. In the case of the $Z^+(4430)$,
since its mass is close to the $D^* D_1$ threshold, it was suggested that it 
could be a $J^P=1^+$ $D^* \bar{D}_1$  molecular state \cite{He:2014nxa} or a cusp
in the $D^*\bar{D}_1$ channel \cite{bugg}. 

The first attempt to demonstrate the resonant behavior of the $Z^+(4430)$ 
state was done by LHCb in Ref.~\cite{Aaij:2014jqa}, where a fit was performed
in which the Breit-Wigner amplitude was replaced by a
combination of independent complex amplitudes at six equally spaced points in 
$m_{\psi(2S)\pi}$ range covering the $Z^+(4430)$ peak 
region. The resulting Argand diagram is consistent with a rapid phase 
transition at the peak of the amplitude, just as expected for a resonance.
 In Ref.~\cite{Chilikin:2014bkk} a similar method was applyed to show the 
resonant behavior of the $Z^+(4200)$.  The 
Breit-Wigner amplitude was replaced by a combination of constant amplitudes, 
with six bins in $m_{\jpsi\pi}$ range covering the $Z^+(4200)$ peak 
and two independent sets of constant amplitudes, to represent the two helicity 
amplitudes of the $Z_c^+(4200)$, $H_0$ and $H_1$. The results in the Argand 
diagram for $H_1$ clearly shows a resonancelike change of the amplitude 
absolute value and phase. However, they argue that because of the Argand 
diagram for the $H_0$ amplitudes has much larger relative errors, it was not 
possible to draw any conclusions from it. In any case, the Argand plot 
approach, proposed by LHCb  experiment, needs a high 
statistics sample to be able to give, in  a undoubted way, the  confirmation 
of the phase variation expected for a regular resonant state. 

In this paper we propose a different method to demonstrate the resonant 
behavior of a state. It is a simple  experimental  method isobar-based 
Amplitude Difference (AD), that can  be used   to extract  the 
phase motion of a complex amplitude in three-body 
heavy-meson decays \cite{ad}. With this method, the  phase variation of a 
generic complex amplitude can be directly revealed through  interference  in 
the Dalitz-plot region where it crosses  a well established  resonant state, 
used as a probe.  This method was successfully applied to Fermilab E791 data 
\cite{UT} to  extract  the   well known  phase motion  of   the scalar 
amplitude $f_0(980)$  observed in  \ds3pi decay. It  was also   successfully 
used to extract   the phase motion, of a resonant scalar amplitude  
$\sigma(500)$ in \d3pi decay \cite{PS2006}, to confirm previous evidences of 
the existence of a light and broad scalar resonance  presented  by  Fermilab 
experiment E791 \cite{prl}.

%\section{Amplitude Difference method} 

In full Dalitz plot  analyses, each possible resonance  amplitude is  
 represented by a Breit-Wigner function multiplied by angular distributions
 associated with the spin of the resonance. 
 The various contributions are combined in a coherent sum, with complex
 coefficients, that are  extracted from  fits to the data.
  The absolute value of the coefficients are related to the relative 
  fraction of each contribution and the phases take into account the final 
  state interaction (FSI) between the resonance and the third particle.

Amplitude Difference method has a different approach. It concentrates in a 
particular region of the Dalitz plot, where the amplitude under study 
crosses a well known resonance amplitude, called  probe amplitude,
 represented by a Breit-Wigner.   The phase variation of the complex amplitude 
can be directly revealed through the interference,  in the Dalitz-plot region, 
where they  cross each other. 

There are two necessary conditions to extract the phase motion  of a generic  
amplitude with the AD method:

\begin{itemize}

\item A crossing region between  the amplitude under 
study and a probe resonance  has to be dominated by these two contributions.

\item The integrated amplitude of the probe resonance must be symmetric 
with respect to an  effective mass squared ($m^2_{eff}$).

\end{itemize}

These two conditions are very well satisfied in many  charmonioum three 
body $B$ decays, where the phase space is large and the charmonium candidates 
are located in the central region of the Dalitz plot, possibly crossing with 
well stablished resonances.   The probe resonances are, in general, placed at  
low $K\pi$, $KK$ or $\pi\pi$ invariant mass. As a consequence, the  charmonioum 
amplitude candidates  must 
cross basically all phase space, or at least the low $K\pi$ region,  to be observed 
by the AD method. Therefore, if the amplitude under study is due to a molecular 
state, this method may not apply. Indeed it was  shown in Ref.~\cite{Bediaga:2012up}
that a loosely bound molecular state can only exist when the relative momentum 
between the two mesons in the molecule is small. This means that this state will 
appear basically only the middle of the Dalitz plot and may not cross the probe 
resonance. This will exclude the direct observation  of the phase variation of
molecular states with the AD method. 

With the two conditions above, we can examine the $B^0\to\psi\pi^-K^+$ decay 
(where $\psi$ represents $\jpsi$ or $\psip$),
and write down an amplitude with two components: 
one representing the probe resonance, $K^*$, placed in the Dalitz variable 
$s_{12}$  
through a Breit-Wigner and the angular distribution, and the other, 
representing the resonance  under study, which we call generically
by $Z$ decaying in $\jpsi\pi$ or $\psip\pi$, placed in the Dalitz variable 
$s_{13}$. 
This simple amplitude must be used only in the small part of the 
phase space where the interference between them occurs. Since the last one 
can have different  dynamical origins, it can be written in a most generic  
way as: 
\beq\label{AZ}
A^Z(s_{13})=\sin \delta (s_{13}) e^{i\delta(s_{13})}\;.
\enq
This  unitary  equation   is able 
to represent amplitudes with slow phase variation, as well as resonances with 
a large phase variation, of the order of $180^0$, around the nominal mass of 
the resonance, in the same way as the Argand plot used by LHCb {\cite{Aaij:2014jqa}}. 
The total amplitude for the
$B^0\to\psi\pi^-K^+$ decay, in the small part of the 
phase space where the interference between the resonances $K^*\to K^+\pi^-$
and $Z^-\to\psi\pi^-$ occurs, 
can be written as \cite{bellez2}:  
\begin{equation}
\left|{\cal A}(s_{12},s_{13})\right|^2 = \sum_{\zeta=1,-1}\left|{A}(s_{12},s_{13},
\zeta)\right|^2,\mbox{ with }{A}(s_{12},s_{13},\zeta)=
%\sum_{\zeta=1,-1}\left|
\sum_{\lambda=-1,0,1}
A^{K^*}_{\lambda\zeta}+\sum_{\lambda'=-1,0,1}
A^{Z}_{\lambda'\zeta}\;,
%\right|^2\;,
\label{A2}
\end{equation}
where $\zeta$, $\lambda$ and $\lambda'$ are the helicities of the lepton pair,
the $K^*$ and $Z$ respectively.
We take the amplitudes of the decays $B^0\to\psi(\to l^+l^-)K^*(\to K^+
\pi^-)$ and $B^0\to K^+Z^-(\to\psi(\to l^+l^-)\pi^-)$ from \cite{bellez2}:
\beqa
A^{K^*}_{\lambda\zeta}&=&H^{K^*}_{\lambda}A^{K^*}(s_{12})d^1_{\lambda0}(
\theta_{K^*})e^{i\lambda\varphi}d^1_{\lambda\zeta}(\theta_{\psi})\;,\nn\\
A^{Z}_{\lambda'\zeta}&=&H^{Z}_{\lambda'}A^{Z}(s_{13})d^1_{0\lambda'}(
\theta_{Z})e^{i\lambda'\tilde{\varphi}}d^1_{\lambda'\zeta}(
\tilde{\theta}_{\psi})e^{i\zeta\alpha}\;.
\label{Akz}
\enqa
In Eqs.~(\ref{Akz}) $H^R_\lambda$ is the helicity amplitude for the decay via
the resonance $R$, $d^J_{mn}(\beta)$ are Wigner $d$ functions, $\theta_R$
is the resonance helicity angle (the angle between $\pi^-$ and $K^+$ or
$\psi$ momenta in the resonance rest frame), $\theta_\psi(\tilde{\theta}_{
\psi})$ is the $\psi$ helicity angle (the angle between $\pi^-$ and $l^-$ 
 momenta in the $\psi$ rest frame), $\varphi(\tilde{\varphi})$ is the angle
between the planes defined by the $(l^+\pi^-)$ and $(K^+\pi^-)$ momenta in 
the $\psi$ rest frame and $\alpha$ is the angle between the planes defined by 
the $(l^+\pi^-)$ and $(l^+K^*)$ momenta in the $\psi$ rest frame. The amplitudes
$H_1^Z$ and $H_{-1}^Z$ are related by parity conservation:
$H_\lambda^Z=H^Z_{-\lambda}$. Finally,
$A^{Z}(s_{13})$ is  given in Eq.~(\ref{AZ}) and  $A^{K^*}(s_{12})$ is
described by a Breit-Wigner:
\beq\label{AK}
A^{K^*}(s_{12})={m_0 \Gamma_0 \over {s_{12} - m^2_0 + im_0\Gamma_0}}\;, 
\enq
where $m_0$ and $\Gamma_0$ are the mass and the width of the $K^*(892)$
respectively.

Let us define
\beqa
a^{K^*}_\zeta e^{i\alpha_\zeta(K^*)}&=&H^{K^*}_{-1}d^1_{-10}(\theta_{K^*})
e^{-i\varphi}d^1_{-1\zeta}(\theta_{\psi})+H^{K^*}_{0}d^1_{00}(\theta_{K^*})
d^1_{0\zeta}(\theta_{\psi})+H^{K^*}_{1}d^1_{10}(\theta_{K^*})
e^{i\varphi}d^1_{1\zeta}(\theta_{\psi}),\nn\\
a^{Z}_\zeta e^{i\alpha_\zeta(Z)}&=&H^{Z}_{1}\left(d^1_{0-1}(\theta_{Z})
e^{-i\tilde{\varphi}}d^1_{-1\zeta}(\tilde{\theta}_{\psi})+d^1_{01}(\theta_{Z})
e^{i\tilde{\varphi}}d^1_{1\zeta}(\tilde{\theta}_{\psi})\right)+H^{Z}_{0}
d^1_{00}(\theta_{Z})d^1_{0\zeta}(\tilde{\theta}_{\psi})\;.
\label{def-al}
\enqa
Using Eqs.~(\ref{Akz}) and (\ref{def-al}), we can write
\begin{equation}
{A}(s_{12},s_{13},\zeta)=\sum_{\lambda=-1,0,1}
A^{K^*}_{\lambda\zeta}+\sum_{\lambda'=-1,0,1}
A^{Z}_{\lambda'\zeta}=A^{K^*}(s_{12})a^{K^*}_\zeta e^{i\alpha_\zeta(K^*)}+
A^Z(s_{13})e^{i\zeta\alpha} a^{Z}_\zeta e^{i\alpha_\zeta(Z)}.
\label{sum-la}
\end{equation}
Therefore,
\beqa
|{A}(s_{12},s_{13},\zeta)|^2 &=& (a^{K^*}_\zeta)^2|A^{K^*}(s_{12})|^2+(
a^{Z}_\zeta)^2\sin\delta(s_{13})+{2a^{K^*}_\zeta a^{Z}_\zeta\sin\delta(s_{13})
m_0\Gamma_0\over(s_{12}-m_0^2)^2+m_0^2\Gamma_0^2}\left[(s_{12}-m_0^2)\cos(
\delta(s_{13})+\beta_\zeta)\right.\nn\\
&-&\left. m_0\Gamma_0\sin(\delta(s_{13})+\beta_\zeta)\right],
\label{A22}
\enqa
where we have used Eqs.~(\ref{AZ}) and (\ref{AK}) and we have defined: 
$\beta_\zeta=\alpha_\zeta(K^*)-\alpha_\zeta(Z)-\zeta\alpha$.

The expression in Eq.~(\ref{A22}) is very similar to the one obtained
in \cite{PS2006}, for two scalar resonances.
For small $\Gamma_0$, $|A^{K^*}(s_{12})|^2$ can be considered as a   
symmetric function, therefore: 
$|A^{K^*}(s_{12} = m_0^2 + \epsilon )|^2- 
  |A^{K^*}(s_{12} = m_0^2 - \epsilon )|^2=0$, where $\epsilon$ is small.
Also, in the small part of the phase space where the interference between the
$K^*$ and the $Z$ occurs, we can suppose that all the angles are almost constant.
Consequently the difference of the amplitudes squared takes the 
simple   form:
\begin{eqnarray}\label{dif}
\mid {A}( m_0^2 + \epsilon, s_{13},\zeta ) \mid^2 
-\mid {A}( m_0^2 - \epsilon, s_{13},\zeta) \mid^2  = 
{4a^{K^*}_\zeta a^{Z}_\zeta m_0\Gamma_0\epsilon\sin\delta(s_{13})
\over\epsilon^2+m_0^2\Gamma_0^2} \cos(\delta(s_{13})+\beta_\zeta)\;.
\enqa
We can rewrite Eq.~(\ref{dif}) as:
\begin{eqnarray}\label{del}
\Delta \mid{A}_\zeta\mid^2 =
\mid {A}( m_0^2 + \epsilon, s_{13},\zeta ) \mid^2 
-\mid {A}( m_0^2 - \epsilon, s_{13},\zeta) \mid^2  = 
 {\cal C_\zeta} ( \sin(2 \delta(s_{13})+ \beta_\zeta) - \sin \beta_\zeta),
 \end{eqnarray} 
\noindent
where ${\cal C_\zeta}=2a^{K^*}_\zeta a^{Z}_\zeta \epsilon m_0 \Gamma_0 /
(\epsilon^2+m_0^2\Gamma^2_0)$. 

Using Eqs.~(\ref{A2}) and (\ref{del}) we can write the difference of the amplitudes 
squared:
\beq\label{eq:sim}
\Delta \mid{\cal A}\mid^2= \mid{\cal A}( m_0^2 + \epsilon, s_{13})\mid^2-
\mid{\cal A}( m_0^2 - \epsilon, s_{13})\mid^2=\sum_{\zeta=-1,1}\Delta \mid{A}_\zeta
\mid^2=\sum_{\zeta=-1,1}{\cal C_\zeta} \left(\sin(2 \delta(s_{13})+ \beta_\zeta) - \sin 
\beta_\zeta\right)\;,
\enq
where $\mid {\cal A}( m_0^2 - \epsilon, s_{13} ) \mid^2$ and 
$\mid {\cal A}( m_0^2 + \epsilon, s_{13}) \mid^2$ are taken from data. 
The above equation can, finally, be rewriten as:
\beq\label{fi}
\Delta \mid{\cal A}\mid^2= \sin(2\delta(s_{13}))\sum_{\zeta=-1,1}{\cal C_\zeta}
\cos\beta_\zeta+\left(\cos(2\delta(s_{13}))-1\right)\sum_{\zeta=-1,1}{\cal C_\zeta}
\sin\beta_\zeta\;.
\enq
$\Delta \mid{\cal A}\mid^2$, in Eq.~(\ref{fi}), directly reflects the 
behavior of $\delta(s_{13})$.   A constant $\Delta\mid{ \cal A}\mid^2$ would 
imply a constant $\delta(s_{13})$, and this would
 be the case of  non-resonant contribution. In the same way, a slow phase 
motion will produce a slowly  varying $\Delta\mid{\cal A}\mid^2$ and a full 
resonance phase motion produces a clear signature in 
$\Delta\mid{\cal A}\mid^2$ with the presence of zero, maximum and 
minimum values.

To clarify these possible  behaviors of $\Delta\mid{\cal A}\mid^2$ and shown the statistic 
feasibility of the AD method, we perform a simple Monte Carlo study. 
To do that we generated two Monte Carlo samples of   $B^0\to \psip \pi^-K^+$ decay 
channel, each one with a sample of $20,000$ events with relative fractions 0.86 and 
0.08 respectively for the $K^*/\psip$ and $Z^-(4430)/K$ contributions,   similarly to the 
observed by the LHCb experiment with 3 fb$^{-1}$ accumulated data~\cite{Aaij:2014jqa}.  
In the first sample the $Z^-(4430)$ enters as a regular resonance with a strong phase 
variation, represented by a Breit-Wigner, while in the second, $Z^-(4430)$ is represented 
by a real bump amplitude with no strong phase associated.  
To simplify this study we use  only one helicity of the lepton pair in 
Eq.(\ref{eq:sim}) and assume that in the little phase space crossing region both  
amplitudes do not have a significant variation due to the angular distribution. 
In both cases we assume zero phase difference between these amplitudes. 
Finally we do not include background components in our simulation.

Fig.~\ref{fig:plots}(a) shows  the $\Delta\mid{\cal A}\mid^2$ distribution for the sample 
with a Breit-Wigner representing the $Z^-(4430)$ particle. 
One can see a clear oscillation behaviour around the zero value of this function, with 
positive and negative regions, placed around the nominal mass value of this charmonium 
state. As it was discussed in ref. \cite{ad}, this particular distribution  is determined
by the phase difference between the two amplitudes. 
Here we assume zero, but any other  possible value to this phase difference produces the 
same signature: positive, negative and zero regions along the crossing region.  
The $\Delta\mid{\cal A}\mid^2$ distribution for the second Monte Carlo sample, 
with no phase variation associated to the $Z^-(4430)$ around the  
$K^{*}$ mass region, is shown in  Fig.~\ref{fig:plots}(b). 
The behaviour are clearly different from Fig.~\ref{fig:plots}(a), with almost  constant 
value for $\Delta\mid{\cal A}\mid^2$ distribution. The mean value of $\Delta\mid{\cal A}
\mid^2$ is shifted from zero due to the constant behaviour  of $\delta(s_{13})$ phase, 
along the  $s_{13}$ variable in the phase space region considered.   

\begin{figure}[tb] 
\begin{overpic}[width=0.49\linewidth]{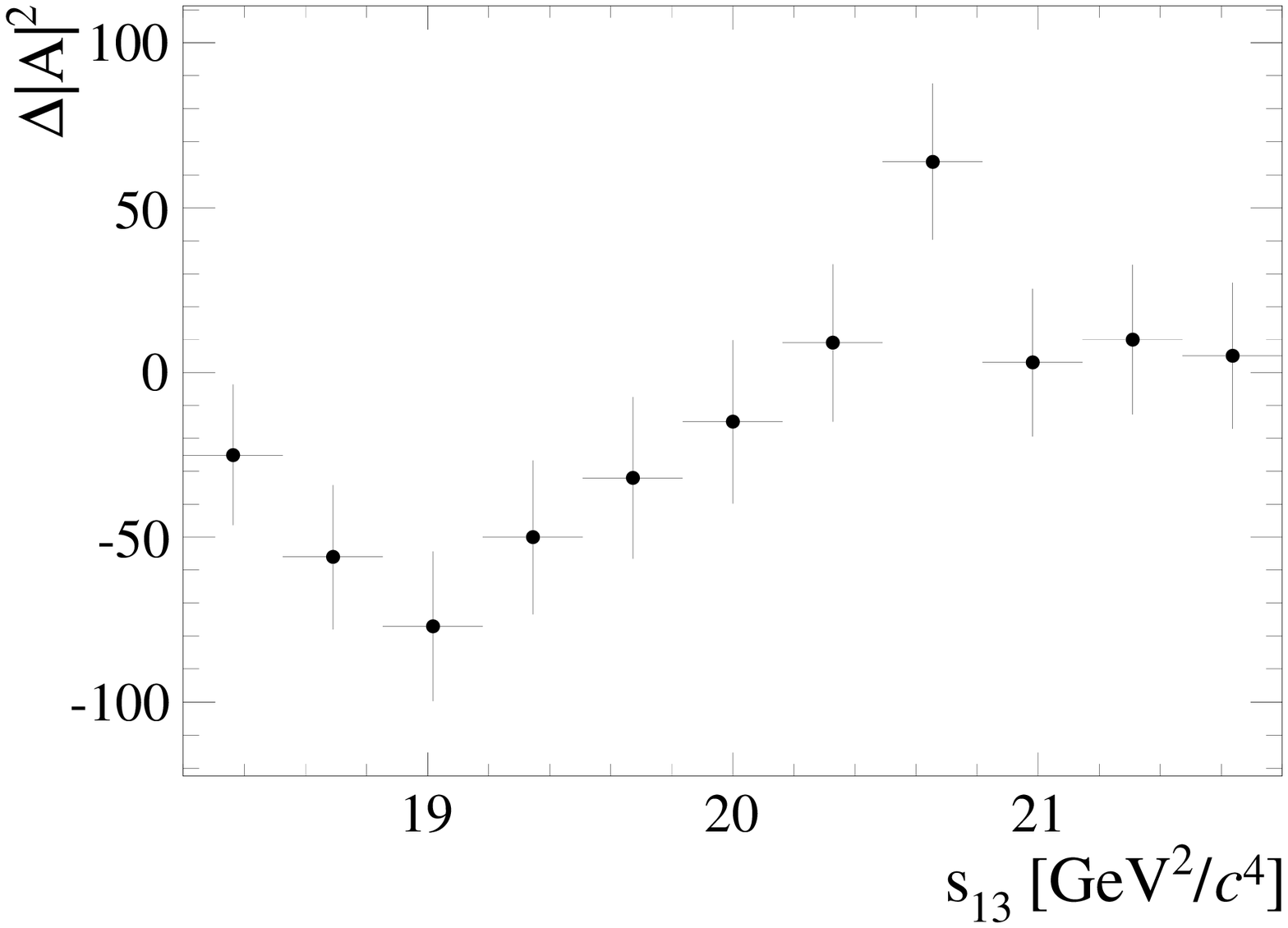}
 \put(20,60){\bf{\Large(a)}}
\end{overpic}
\begin{overpic}[width=0.49\linewidth]{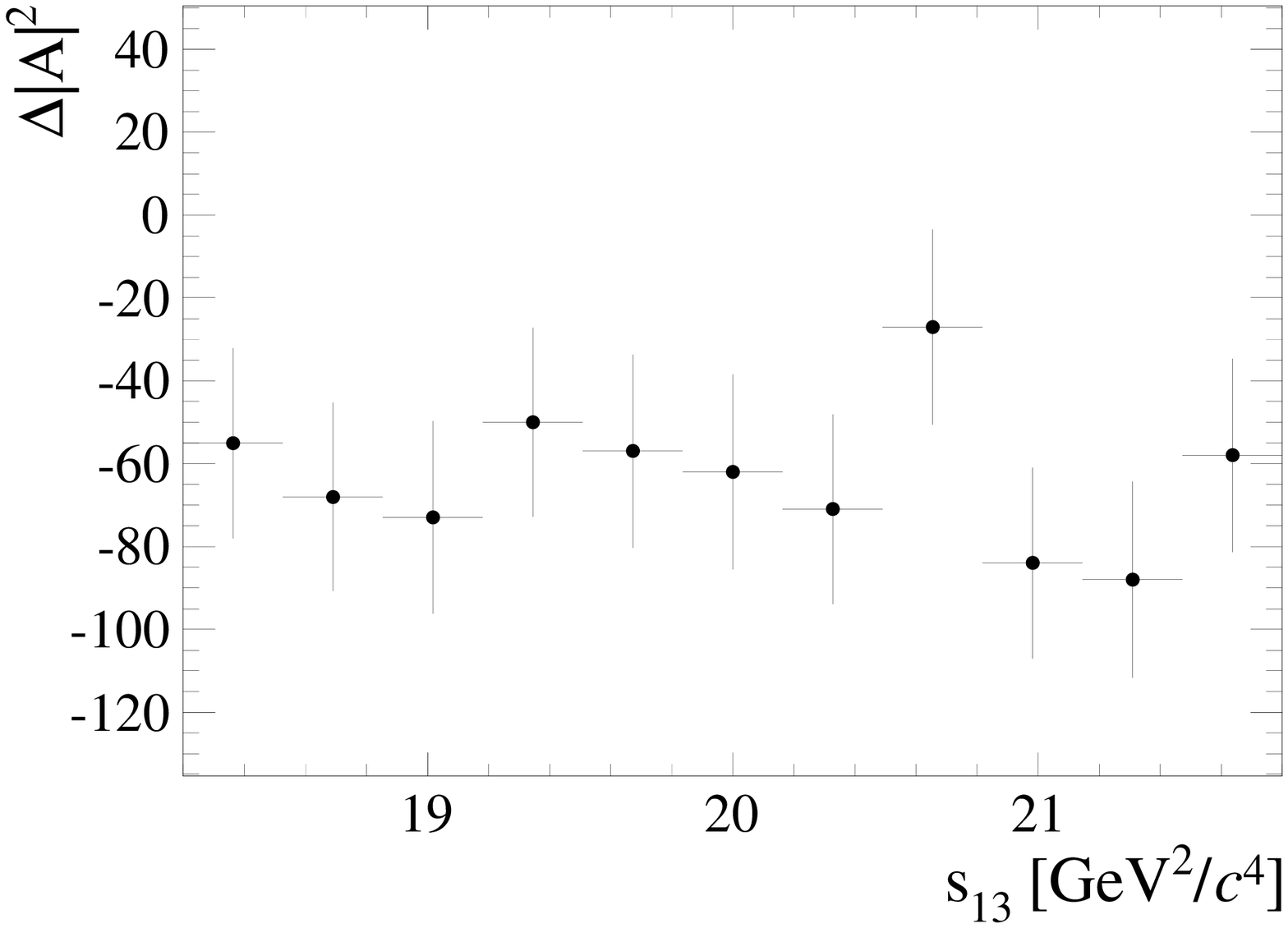}
 \put(20,60){\bf{\Large(b)}}
\end{overpic}
\caption{$\Delta \mid{\cal A}\mid^2$ $= (\mid{\cal A}( m_0^2 + \epsilon, s_{13})\mid^2-
\mid{\cal A}( m_0^2 - \epsilon, s_{13})\mid^2)$, as a function of $s_{13}=m^2(\psip\pi)$, 
for a sample of 20,000 signal events, using $K^{*}$ as a probe with $m_0=0.89594$ and 
$\epsilon=0.06$. Two cases are illustrated: (a) considering both resonances represented 
as Breit-Wigner, and (b) representing the $K^{*}$ resonance as a Breit-Wigner and a real 
bump  around $s_{13}=20$ GeV$^2/c^4$ as a non-resonant component.
}
\label{fig:plots} 
\end{figure}  

In conclusion, we have discussed the new findings of several experiments operating during 
the last decade, with many indications of new charmonium states. We have indetified the 
need to have a direct confirmation of these states through the study of the phase 
variation associated to a real resonant state. The first attempt to demonstrate such
resonant behavior was done by the LHCb Collaboration \cite{Aaij:2014jqa}, for the charged 
charmonioum state $Z^-(4430)$ observed  in the  $B^0\to \psip \pi^-K^+$ decay. 
In this paper we present an alternative method, called isobar-based Amplitude Difference
(AD), already used in charm three body decays \cite{ad,UT,PS2006}, that can be used in 
cases where the amplitude under study crosses, in the Dalitz plot, a well established 
resonance. We have tested the method for the $Z^-(4430)$ state by generating two Monte 
Carlo samples of   $B^0\to \psip \pi^-K^+$ decay channel. The first where the 
$Z^-(4430)$ is considered as a regular resonance with a strong phase variation, 
represented by a Breit-Wigner, and the second where the $Z^-(4430)$ is represented by a 
real bump amplitude with no strong phase associated and, therefore, is not a real 
particle.  Each one of these Monte Carlo simulations were generated with  
samples  similar to the observed by the LHCb experiment ~\cite{Aaij:2014jqa}. 
We have shown that, in the first case, the AD method gives a clear oscillation 
behaviour related to the phase variation associated to a real resonant state. 
For the second case no oscillation behaviour was observed.

As a final remark we would like to stress that the AD method  can be used in a 
complementary way to the LHCb approach, to determine the internal structure of  
$Z^-(4430)$. By construction the AD method can only be used when the amplitude under 
study crosses, in the Dalitz plot, a well stablished resonance, which is considered as 
the probe resonance. In the case under consideration, the $Z^-(4430)$, the probe 
resonance is  the $K^{*}$, which will be located at low  $\pi^-K^+$ invariant  mass 
in the  dalitz plot for the $B^0\to \psip \pi^-K^+$ decay. If the $Z^-(4430)$ is a 
loosely bound molecular state, as suggested in \cite{He:2014nxa,Branz:2010sh}, it will 
only appear in the central region of the Dalitz plot for the three body phase space,
as discussed in Ref. \cite{Bediaga:2012up}, and will not cross the probe resonance.
Therefore, since the LHCb Collaboration have already shown the resonant behaviour
of the $Z^-(4430)$ through the Argand diagram \cite{Aaij:2014jqa}, if no phase variation
is observed with the proposed AD method, this could be used to confirm the molecular
structure of the  $Z^-(4430)$. By the other hand, if the phase variation is also observed
with the AD method, this would confirm the tetraquark structure for the  $Z^-(4430)$.
We strongly suggest the LHCb Collaboration to perform such analysis.

\section*{Acknowledgment} 
 
\noindent  
 This work has been partially supported by FAPESP and CNPq.

\end{document}